\documentclass[aps,pre,preprint, superscriptaddress]{revtex4}
\usepackage{latexsym}
\usepackage{natbib}
\usepackage{amsmath}
\usepackage{amsfonts}
\usepackage{amssymb}
\usepackage[T1]{fontenc}
\usepackage[dvips]{color,graphicx}   
\usepackage{verbatim}   
\usepackage{t1enc}
\usepackage{anysize}
\usepackage{tocbibind}
\usepackage{subfig}
\usepackage[percent]{overpic} 
\usepackage{float}  
\usepackage{appendix} 
\usepackage{booktabs} 
\usepackage{natbib}
\usepackage{multirow}
\usepackage{array}
\usepackage[alpine,clock,electronic,geometry,misc,weather]{ifsym}

\begin{document}

\title{Temporal percolation of the susceptible network in an
    epidemic spreading} 
\author{Valdez L. D.}
\email{ldvaldes[at]mdp.edu.ar}\affiliation{Instituto de
  Investigaciones F\'isicas de Mar del Plata (IFIMAR)-Departamento de
  F\'isica, Facultad de Ciencias Exactas y Naturales, Universidad
  Nacional de Mar del Plata-CONICET, Funes 3350, (7600) Mar del Plata,
  Argentina.}  \author{Macri P. A.} \affiliation{Instituto de
  Investigaciones F\'isicas de Mar del Plata (IFIMAR)-Departamento de
  F\'isica, Facultad de Ciencias Exactas y Naturales, Universidad
  Nacional de Mar del Plata-CONICET, Funes 3350, (7600) Mar del Plata,
  Argentina.}  \author{Braunstein L. A.}  \affiliation{Instituto de
  Investigaciones F\'isicas de Mar del Plata (IFIMAR)-Departamento de
  F\'isica, Facultad de Ciencias Exactas y Naturales, Universidad
  Nacional de Mar del Plata-CONICET, Funes 3350, (7600) Mar del Plata,
  Argentina.}  \affiliation{Center for Polymer Studies, Boston
  University, CPS, 590 Commonwealth Av, Boston, Massachusetts 02215,
  USA}

\begin{abstract}
In this work, we study the evolution of the susceptible individuals
during the spread of an epidemic modeled by the
susceptible-infected-recovered (SIR) process spreading on the top of
complex networks. Using an edge-based compartmental approach and
percolation tools, we find that a time-dependent quantity $\Phi_S(t)$,
namely, the probability that a given neighbor of a node is susceptible
at time $t$, is the control parameter of a node void percolation
process involving those nodes on the network not-reached by the
disease. We show that there exists a critical time $t_c$ above which
the giant susceptible component is destroyed. As a consequence, in
order to preserve a macroscopic connected fraction of the network
composed by healthy individuals which guarantee its functionality, any
mitigation strategy should be implemented before this critical time
$t_c$. Our theoretical results are confirmed by extensive simulations
of the SIR process.
\end{abstract}
\maketitle
\section{Introduction}

The study of epidemic spreading has been one of the most successful
applications on networks science. Recent outbreaks of new influenza
strains like the H1N1~\cite{Baj_01} and the H5N5 flu or the Severe
Acute Respiratory Syndrome (SARS)~\cite{Col_01}, which are
characterized by a high rate of mortality and/or fast propagation
velocity, motivate the development of epidemic models that capture the
main features of the spread of those diseases. In particular,
mathematical tools applied to model epidemics are very important since
they allow to understand how a disease impact on the society, helping
to develop new policies to slow down its spreading.

One of the simplest models that reproduce seasonal diseases, such as
influenza, is the susceptible-infected-recovered (SIR)
model~\cite{Boc_01,And_01}, which has been the subject of extensive
theoretical and numerical research on complex
networks~\cite{Boc_01}. In the SIR model the individuals can be in one
of three states, susceptible, infected or recovered. In its discrete
formulation~\cite{Lag_02,New_05,Par_01}, at each time step, infected
individuals infect their susceptible neighbors with probability
$\beta$ and recover at a fixed time $t_r$ since they were infected,
called recovery time. According to these rules, the disease spreads on
the contact network until it reaches the steady state where there are
only susceptible and recovered individuals. It was found that the
steady state of the SIR model can be mapped into a link percolation
problem which provides a theoretical framework to study this
process~\cite{Gra_01,New_05,Mil_02,Ken_01}. It is known that the size
of the infection, defined as the fraction of recovered individuals at
the steady state, is governed by the effective probability of
infection or transmissibility $T$ of the disease which depends on
$\beta$ and $t_r$. In the SIR model, the size of the infection is the
order parameter of a second order phase transition with a critical
threshold transmissibility $T_c$. Below $T_c$ the disease is an
outbreak, where the infection reaches a small fraction of the
population while above $T_{c}$ an epidemic develops exactly as in a
link percolation process~\cite{Gra_01,New_05,Mil_02,Ken_01}. In
uncorrelated infinite networks this threshold is given by
$T_{c}=1/(\kappa-1)$~\cite{Coh_01,New_05}, where $\kappa=\langle
k^2\rangle/\langle k \rangle$ is the branching factor of the network,
and $\langle k \rangle$ and $\langle k^2\rangle$ are the first and the
second moment, respectively, of the degree distribution $P (k)$. Here,
$k$ is the degree or number of links that a node can have with
$k_{min}\leq k \leq k_{max}$. For Erd\"os-R\'enyi networks (ER), the
degree distribution is $P(k)=e^{-\langle k \rangle}\langle k
\rangle^{k}/k!$ and the threshold is found at $T_{c}=1/\langle k
\rangle$. However, most of the real networks have a heterogeneous
degree distribution that is better represented by a pure Scale-Free
network (SF) with $P (k) \sim k ^{-\lambda}$, where $\lambda$
measures the broadness of the distribution. In the thermodynamic
limit, for SF networks with $2<\lambda <3$, $\langle k^2\rangle \to
\infty$ and as a consequence, the critical transmissibility $T_c\to 0$
which means that the epidemic spreads for any value of
$T$~\cite{New_05,Coh_01}. However, due to finite size effects, real
networks have finite critical transmisibilities.

In a recent paper, using a generating function formalism,
Newman~\cite{New_06} showed that at the steady state of the SIR model
there exists a second threshold $T^{*}$ above which the residual
network composed by the biggest giant susceptible cluster that
remains after a first propagation, is destroyed. From an
epidemiological point of view, this implies that if a disease
spreads for a second time on the residual network, it cannot become an
epidemic. On the other hand, Valdez $et\;al.$~\cite{Val_01} showed
that $T^{*}$ is an important parameter to determine the efficiency of
a mitigation or control strategy, because any strategy that decrease
the transmissibility below $T^{*}$, can protect a large and connected
cluster of susceptible individuals. Using a percolation framework,
they explained the lost of the susceptible giant cluster as a
not-random node percolation process, that they called node void
percolation, in which a susceptible individual corresponds to a void
node in link percolation.

Even though percolation theory was very useful to describe the steady
state of the SIR model on complex networks, it is still very
challenging to explain the dynamics of the model to develop
intervention strategies before the epidemic spreads to a large
fraction of the population. To describe the dynamics of epidemic
spreading on networks, recently some researchers developed
differential rate equations for the SIR model that take into account
the network topology. Lindquist $et\;al.$~\cite{Lin_01} introduced an
``effective degree'' approach through a large system of ordinary
differential equations. Under this approach, the nodes and their
neighbors are categorized by their disease state (susceptible,
infected, recovered) and each differential equation compute the
evolution of the fraction of susceptible or infected nodes with a
number $i$ and $s$ of infected and susceptible neighbors,
respectively, with $0\leq i\leq k_{max}$ and $0\leq s\leq k_{max}$. As
a result, a system with $\mathcal{O} \left(k_{max}^{2}\right)$
equations needs to be solved. This approach represents accurately the
evolution of the number of infected individuals, but at a high
computational cost. On the other hand, Miller~\cite{Mil_03} and Miller
$et\;al.$~\cite{Mil_04,Vol_01} proposed an ingenious approach to
describe the evolution of a SIR process with rates by means of an
edge-based compartmental model (EBCM)~\cite{Mil_03,Mil_04} which has
the advantage to describe the dynamical spreading of an epidemic with
only a few equations. With these equations, the authors found accurate
results for the evolution of the number of infected individuals for
static and dynamic evolutive topologies like ``edge swapping'' and
``dormant contacts'' for transmissibilities above the critical
threshold~\cite{Mil_04}.

While most of the literature is focused on studying the evolution of
the fraction of infected or susceptible individuals, it has not yet
been investigated how the epidemic spread affects the evolution of the
network composed by the susceptible individuals. Understanding this
problem is important because the network composed by the healthy
individuals is the network that sustains the functionality of a
society, e.g. the economy of a region. In this paper we present a
novel idea for the SIR model, based on a dynamical study of the
network composed by susceptible individuals. We show that the temporal
decreasing of the size of the giant susceptible cluster can be
described as a dynamic void node percolation process with an
instantaneous void control parameter. We find that there exists a
critical time $t_c$ above which the giant susceptible component
overcomes a temporal second order phase transition with mean field
exponents. The paper is organized as following: in Methods and Results
we present the theoretical framework to derive the evolution
equations. Then we study the evolution of the giant susceptible
cluster and its temporal critical behavior. Finally we present our
conclusions.

\section{Methods and Results}

\subsection*{Theoretical framework}\label{Sec_Inf_Cur}
The evolution equations of the dynamic SIR model provide the basis
for analyzing theoretically novel magnitudes that could be useful for
epidemiologists and authorities to plan policies to stop a disease
before an epidemic develops. In the SIR model, initially, all the
nodes are susceptible except for one node randomly infected, that
represents the index case from which the disease spreads. The infected
individual transmits the disease to susceptible neighbors with
probability $\beta$ and recovers $t_{r}$ time units since he was
infected. For the SIR with fixed recovery time, the transmissibility
is given by $T(\beta,t_r)\equiv T=1-(1-\beta)^{t_{r}}$~\cite{Val_01}.

In order to study the evolution of the states of the individuals in
the SIR with fixed recovery time, we use the edge-based compartmental
model (EBCM)~\cite{Vol_01,Mil_03,Mil_04}. The EBCM is based on a
generating function formalism, widely implemented in branching and
percolation process on complex
networks~\cite{Dor_03,Boc_01,New_03,New_07}. For a branching process
that spreads on uncorrelated networks, such as the tree of infected
individuals, two generating functions that contain the information of
the topology of these networks are defined. The first one is the
generating function of the node degree distribution $P(k)$ which is
given by $G_{0}(x)=\sum_{k}P(k)x^{k}$. The second one is the
generating function of the degree distribution of the first neighbors
of a node, also called excess degree distribution $P_1(k)\equiv
kP(k)/\langle k \rangle$, given by $G_{1}(x)=\sum_{k}kP(k)/\langle
k \rangle x^{k-1}$. Here, $P_1(k)$ is the probability to
reach a neighbor of a node, following a link. It is straightforward
that the mean connectivity of the nodes is $\langle k \rangle
=G_{0}^{'}(1)$.

Denoting the fraction of susceptible, infected and recovered
individuals at time $t$ by $S(t)$, $I(t)$ and $R(t)$, respectively,
the EBCM approach describes the evolution of the probability that a
node (which we call root node) is susceptible.  In order to compute
this probability, an edge is randomly chosen and a direction is given,
in which the node in the target of the arrow is the root, and the base
is its neighbor. Disallowing that the root infects the neighbor,
$\theta(t)\equiv \theta_{t}$ is the probability that the neighbor does
not transmit the disease to the root, with $\theta_t$ given by
\begin{eqnarray}\label{thetaEqPhi}
\theta_t=\Phi_{S}(t)+\Phi_{I}(t)+\Phi_{R}(t),
\end{eqnarray}
where $\Phi_{S}(t)$, $\Phi_{R}(t)$ and $\Phi_{I}(t)$ are the
probabilities that the neighbor is susceptible, recovered, or infected
but has not transmitted yet the disease to the root. The probability
that a root node with connectivity $k$ is susceptible is therefore
$\theta_{t}^{k}$ and the fraction of susceptible nodes is
$S(t)=\sum_{k} P(k)\theta_{t}^{k}=G_{0}(\theta_{t})$. This approach
simplifies the calculations, reducing the problem to finding an
evolution equation for $\theta_t$, from where the evolution of $S(t)$,
$R(t)$ and $I(t)$ is derived.  Thus, using the EBCM approach adapted
to SIR with fixed $t_r$ (see Appendix Sec.I), the
evolutions of $\theta_t$, $\Phi_{S}(t)$ and $\Phi_{I}(t)$ are given by the
deterministic equations

\begin{eqnarray}
\Delta{\theta_t}&=&-\beta\Phi_{I}(t),\label{delta_theta}\\
\Delta{\Phi}_{S}(t)&=&G_{1}(\theta_{t+1})-G_{1}(\theta_{t}),\label{deltaPhiS}\\
\Delta{\Phi}_{I}(t)&=&-\beta\Phi_{I}(t)-\Delta{\Phi}_{S}(t)+(1-T)\Delta{\Phi}_{S}(t-t_r),\label{deltaPhi}
\end{eqnarray}
where $\Delta$ is the discrete change of the variables between times
$t$ and $t+1$. Eq.~(\ref{delta_theta}) represents the decrease of
$\theta_t$ when a infected neighbor transmits the
disease. Eq.~(\ref{deltaPhiS}) represents the decrease of
$\Phi_{S}(t)$ when a susceptible neighbor is infected (notice that
$\Delta \Phi_S(t)<0$). This term contributes to an increase of
$\Phi_{I}(t)$ in Eq.~(\ref{deltaPhi}) where the first term represents the
decrease of $\Phi_{I}(t)$ when the links transmit the disease, the second
term corresponds to the term of Eq.~(\ref{deltaPhiS}) mentioned above
and the third term represents the decrease of $\Phi_{I}(t)$ due to the
recovery of infected individuals.

From the above equations, the evolution of the fraction of infected
individuals can be computed as
\begin{eqnarray}\label{EqInfec}
\Delta{I}(t)&=& -\Delta S(t)+\Delta S(t-t_r),
\end{eqnarray}
where the first term represents the fraction of new infected
individuals (see Appendix Sec.I). The second term
represents the recovery of infected individuals that have been
infected $t_{r}$ time units ago.

These difference equations correctly describe the evolution of $S(t)$,
$I(t)$ and $R(t)$ above the criticallity for all values of $t_r$ and
$\beta$ (see Appendix Sec.I). In the next section, we will
show that combining this approach and dynamic percolation, we can
describe the time-dependent evolution of the susceptible individuals
in the SIR model as a dynamic void node percolation process for any
value of $t_r$.

\subsection{Temporal percolation of susceptible individuals}\label{Sec_EvolSusCl}
In Ref.~\cite{Val_01} it was found that the process under which the
susceptible clusters size decrease can be explained with node void
percolation defined below that as we will show can be related with the
dynamic SIR process.

In the steady state of the SIR model an epidemic cluster is equivalent
to a Leath growth process~\cite{Lea_01,Bra_01} with a link occupancy
probability $T$. The Leath process on complex networks generates a
single cluster that represents the infection tree for a given value of
the transmission probability $T$. Denoting by $f_{n}(T)$ the
probability that a cluster reaches the $nth$ generation following a
link, the probability $f_{\infty}(T)$ that a link leads to a giant
component ($n\to \infty$) is given by~\cite{Bra_01,Val_01}
\begin{equation}\label{it03}
f_{\infty}(T)=1-\sum_{k=1}^{\infty}\frac{kP(k)}{\langle k\rangle} \left[ 1-T\;f_{\infty}(T)\right]^{k-1},
\end{equation}
where $f_{\infty}(T)$ is the solution of
\begin{equation}\label{eqpinf}
f_{\infty}(T)=1-G_{1}\left[1-T\;f_{\infty}(T)\right].
\end{equation}
As the ``infectious'' cluster grows from a root, generation by
generation, the sizes of the void clusters, $i.e.$ the nodes not
reached by the disease, are reduced as in a node dilution process,
since when a link is traversed a void cluster loses a node and all
its edges. As a consequence, for large generations $f_{\infty}(T)$ can
also be interpreted as the probability that a void cluster loses a
node. However, in this kind of percolation process the void nodes are
not killed at random, instead they are removed following a link. We
call this type of percolation ``node void percolation''. If we denote
by $1-V^{s}$ the probability that a void node is removed due to the
occupancy of a link, at the steady state the following relation holds
\begin{eqnarray}
1-V^{s}=f_{\infty}(T).
\end{eqnarray}
Then $V^{s}$ is the probability that a void node is not removed due to
the fact that the link has not been traversed. Thus, $V^{s}$ is
equivalent to $\Phi_{S}(t\to \infty)$ because the void nodes
correspond to the susceptible individuals in the steady state. As in
any percolation process, there is a critical probability $V^{s}_c$ at
which the void network undergoes a second order phase
transition. Above $V^{s}_c$ a giant void component exist while at and
below $V^{s}_c$ void nodes belong only to finite components. In
epidemic terms, this means that at $V_c^{s}$ only finite susceptible
clusters can be reached. As a consequence, the fraction of links
$T^{*}$ needed to reach this point fulfills~\cite{Val_01}
\begin{equation}\label{condition_p}
V_{c}^{s}=1-f_{\infty}(T^{*}).
\end{equation}
Therefore, from Eqs.~(\ref{eqpinf})~and~(\ref{condition_p}) we obtain
\begin{equation}\label{pestrellaFinal}
V_{c}^{s}=G_{1}\left[1-T^{*}(1-V_{c}^{s})\right],
\end{equation}
where $T^{*}$ is the solution of Eq.~(\ref{pestrellaFinal}). This
result shows that at the steady state, for $T\geq T^{*}$, we have
$V^{s}<V^{s}_c$ and therefore the size of the giant susceptible
cluster $S_1\to0$~\cite{Val_01}. Even though static percolation is a
useful tool to analyze the final size of the giant component of
susceptible individuals~\cite{New_06}, it is very important to know
the evolution of $S_1(t)$, since it can be used as a criteria to begin
or to increase an intervention to protect a large fraction of the
susceptible population~\cite{Val_01}. As we will show below, $S_1(t)$
can be fully related with a node void percolation process at every
instant $t$.

\begin{figure}[H]
\centering
\vspace{0.3cm}
\includegraphics[scale=0.28]{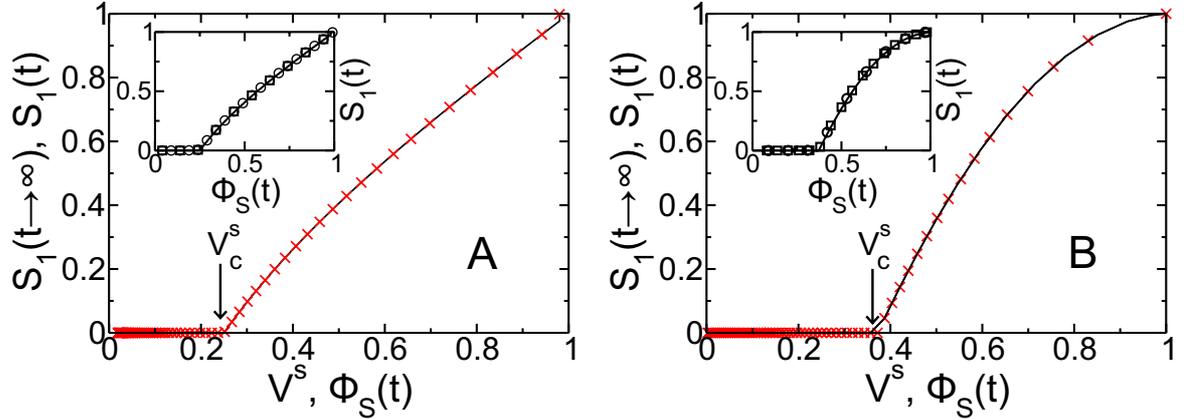}
\vspace{0.5cm}
\caption{Equivalence between $\Phi_{S}(t)$ and $V^s$. $S_{1}(t\to
  \infty)$ as a function of $V^s$ ($\times$) obtained in
  Refs.~\cite{New_06,Val_01} and $S_{1}(t)$ as a function of
  $\Phi_{S}(t)$ (solid line) obtained from
  Eqs.~(\ref{deltaPhiS})-(\ref{delta_theta}) and
  (\ref{EqSusTotal1})-(\ref{EqSusTotal2}) with $N=10^5$ and mean
  connectivity 4.07 in the giant component for (A) a ER network with
  $\langle k \rangle =4$ and (B) SF network with $\lambda=2.63$,
  $k_{min}=2$ and $\langle k \rangle =4.07$. In the insets we show
  $S_1(t)$ as a function of $\Phi_S(t)$ from the simulations (symbols)
  and from Eqs.~(\ref{deltaPhiS})-(\ref{delta_theta}) and
  (\ref{EqSusTotal1})-(\ref{EqSusTotal2}) (solid line) for $t_r=1$
  ($\square$) and $t_r=20$ ($\bigcirc$). (Color
  online).}\label{EquicConcepts}
\end{figure}

In order to describe the evolution of the size of the giant
susceptible cluster, we define $\omega_t$ as the probability that a
neighbor of a root not connected to the giant susceptible cluster has
not yet transmitted the disease to the root at time $t$. This is
possible if the neighbor of the root node is infected but has not yet
transmitted the disease, recovered or susceptible but not connected to
the giant susceptible cluster, with probabilities $\Phi_{I}(t)$,
$\Phi_{R}(t)$ and $G_{1}(\omega_t)$ respectively. Similarly to
$\theta_t$ (see Eq.~(\ref{thetaEqPhi})), these probabilities satisfy
the relation
\begin{eqnarray}\label{Friday}
\Phi_{R}(t)+\Phi_{I}(t)+G_{1}(\omega_t)=\omega_t,
\end{eqnarray}
where $G_{1}(\omega_{t})$ is the generating function of the neighbor
of a root not connected to the giant susceptible cluster. From
Eq.~(\ref{thetaEqPhi}),
$\Phi_{I}(t)+\Phi_{R}(t)=\theta_t-\Phi_{S}(t)=\theta_t-G_{1}(\theta_t)$. Then
Eq.~(\ref{Friday}) can be rewritten as,
\begin{eqnarray}
\omega_{t}-G_{1}(\omega_{t})&=&\theta_{t}-G_{1}(\theta_{t}),\label{EqSusTotal1}
\end{eqnarray}
and the evolution of $S_{1}(t)$ is given by
\begin{eqnarray}
S_{1}(t)&=&G_{0}(\theta_{t})-G_{0}(\omega_{t}),\label{EqSusTotal2}
\end{eqnarray}
where $G_{0}(\theta_t)$ is the total fraction of susceptible
individuals and $G_{0}(\omega_t)$ is the fraction of individuals
belonging to finite susceptible clusters at time $t$. Notice that the
dynamical Eqs.~(\ref{EqSusTotal1})~and~(\ref{EqSusTotal2}) are a
time-dependent versions of the ones derived in Ref.~\cite{New_06} for
the steady state ($t \to \infty$) of the SIR model. This suggests that
the evolution of the giant susceptible or percolating void cluster can
be thought as a temporal percolation process. Thus, the magnitudes
derived for the static percolation of the susceptible individuals have
a dynamical counterpart. As a result, $V^{s}$ and $\Phi_{S}(t)$, are
equivalent not only at the steady state, but also at every instant of
time. In order to show the equivalence, in Fig.~\ref{EquicConcepts} we
show in the same plot $S_1(t)$ as a function of $\Phi_{S}(t)$,
obtained from Eqs.~(\ref{deltaPhiS})-(\ref{delta_theta}) and
(\ref{EqSusTotal1})-(\ref{EqSusTotal2}), and the steady state
$S_{1}(t \to \infty)$ as a function of $V^s$~\cite{New_06} for ER and
SF networks with the same $\langle k \rangle$ and $N$ for
$T=0.76>T^*$.

As we can see, the static curve $S_1(t\to \infty)$ as a function of
$V^s$ is the same as $S_1(t)$ as a function of $\Phi_{S}(t)$ and they
coincide with the simulations for different values of $t_r$ which
shows the equivalence between $V^s$ and $\Phi_S(t)$ at every instant
of time and not only at the steady state (for details of the
simulations see Appendix Sec.I). Thus our process can be
explained by a dynamic percolation with an instantaneous void
transmissibility $V^s\equiv \Phi_S(t)$.

\begin{figure}[H]
\centering
\vspace{0.3cm}
 \includegraphics[scale=0.4]{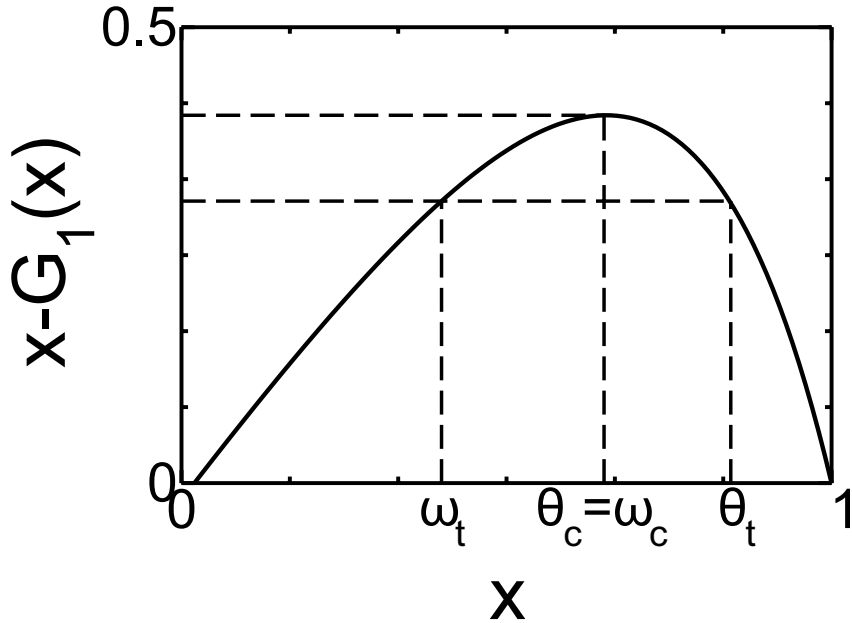}
  \vspace{0.5cm}
\caption{Schematic of the behavior of Eq.~(\ref{EqSusTotal1}) for
  $T>T^{*}$. From the initial condition $\theta_0=\theta(t=0)=1$,
  $\theta_{t}$ and $\omega_{t}$, satisfies
  Eq.~(\ref{EqSusTotal1}). For $\theta_{t}\neq \omega_{t}$ we have two
  solutions that correspond to $S_1(t)>0$. When $\theta_{t}$ reaches
  the maximum of the function $x-G_{1}(x)$, $\theta_{c}=\omega_{c}$,
  the giant susceptible component is destroyed. The dashed lines are
  used as a guide to show the possible solutions of
  Eq.~(\ref{EqSusTotal1}).}\label{Graf}
\end{figure}

With our theoretical formulation, we will show that there is a
critical time $t_c$ at which the giant susceptible cluster disappears
that correspond to the time at which $\Phi_{S}(t_{c})=V^{s}_{c}$. In
order to prove this, notice that according to Eq.~(\ref{EqSusTotal1}),
$\theta_{t}$ and $\omega_{t}$ can be thought as two points with the
same image of the function $x-G_{1}(x)$. Solving this equation for the
variable $\omega_{t}$ above $T^*$, two solutions are found since the
curve $x-G_{1}(x)$ is a concave function for $x>0$ as can be seen in
Fig.~\ref{Graf}. One of the solutions is the trivial one, for which
$S_1(t)=0$, that corresponds to the maximum of the function $x-G(x)$
at $\theta_{t_c}=\omega_{t_c}\equiv \omega_{c}$. Then the giant
susceptible cluster is destroyed at the point $\omega_{c}$ which
fulfills
\begin{eqnarray}\label{Eqmax1}
\left[x-G_{1}(x)\right]^{'}\big|_{w_c}=0,
\end{eqnarray}
then,
\begin{eqnarray}
w_{c}=\left(G_{1}^{'}\right)^{-1}(1).
\end{eqnarray}
Thus when Eq.~(\ref{Eqmax1}) is satisfied, the giant susceptible
cluster disappears and $\Phi_{S}(t_{c})\equiv V_{c}^s=
G_1(\omega_c=\theta_{c})=G_{1}\left[\left(G_{1}^{'}\right)^{-1}(1)\right]$,
$i.e.$
\begin{eqnarray}\label{Eqmax33}
\Phi_{S}(t_{c})=G_{1}\left[\left(G_{1}^{'}\right)^{-1}(1)\right].  
\end{eqnarray}
For ER networks it is straightforward to show that
$\Phi_{S}(t_{c})=1/\langle k \rangle$.

\begin{figure}[H]
\centering
\vspace{1.0cm}
\includegraphics[scale=0.28]{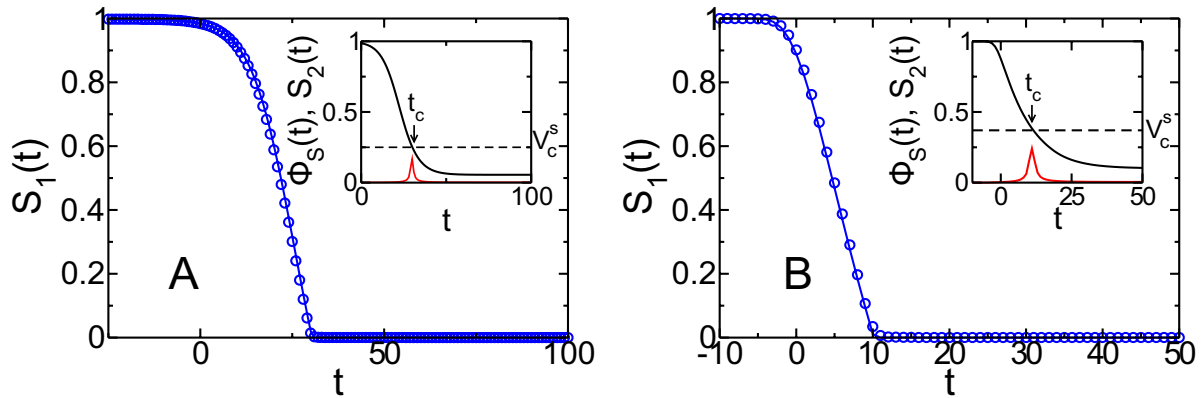}
\vspace{0.5cm}
\caption{Time evolution of $S_1(t)$ for $t_r=20$ and $\beta=0.07$
  ($T=0.76$) and mean connectivity $4.07$ in the giant component for
  (A) a ER network with $\langle k \rangle=4$ ($T^{*}=0.46$) and (B) a
  SF networks with $\lambda=2.63$, minimal connectivity $k_{min}=2$
  and $\langle k \rangle =4.07$ ($T^{*}=0.38$). The symbols correspond
  to the simulations with the time shifted to $t=0$ when $1$\% of the
  individuals are infected, and the solid lines correspond to the
  theoretical solutions $S_1(t)$ (blue solid line) of
  Eqs.~(\ref{EqSusTotal1})-(\ref{EqSusTotal2}). In the insets we show
  the size of the second biggest susceptible cluster $S_2(t)$ (red
  solid line) and the evolution of $\Phi_S(t)$ (black solid line)
  obtained from simulations. The value of $\Phi_S(t_c)=V^s_{c}$
  (dashed line) was obtained from Eq.~(\ref{Eqmax33}). $S_2(t)$ has
  been amplified by a factor of 50 in order to show it on the same
  scale as the rest of the curves. The simulations are averaged over
  1000 network realizations with $N=10^5$. (Color
  online).}\label{GC_Sus_Analyt}
\end{figure}

In Fig.~\ref{GC_Sus_Analyt} we plot the time evolution of the fraction
of susceptible individuals $S_{1}(t)$ in the susceptible giant
component as a function of $t$ for ER and SF networks obtained from
the theory and the simulations, for a transmissibility $T$ above
$T^{*}$.

As shown in Fig.~\ref{GC_Sus_Analyt}, there is an excellent agreement
between the theoretical curve $S_{1}(t)$, obtained from
Eqs.~(\ref{EqSusTotal1}) and (\ref{EqSusTotal2}), and the simulations
which validate that percolation tools can be used to describe the time
dependence of the susceptible individuals in the SIR process for
$T>T^{*}$. On the other hand, in the figure we can see that for
$T>T^{*}$, the giant susceptible cluster $S_{1}(t)$ is destroyed at
$t=t_c$ which occurs exactly at $\Phi_S(t_c)=V^{s}_c$ (see the insets
of Fig.~\ref{GC_Sus_Analyt}). Our results show that $\Phi_S(t)$ can be
used to determine whether a giant susceptible cluster exists at a
given time. In turn, in the insets of Fig.~\ref{GC_Sus_Analyt} we can
see that the size of the second susceptible cluster $S_2(t)$ has a
sharp peak around $t_{c}$, indicating that, as in static percolation,
the susceptible individuals overcome a second order phase
transition. However, this transition is not given by a random node
percolation process. As the disease spreads through the links, the
susceptible individuals are removed with probability proportional to
$kP(k)$, $i.e.$, the susceptible network loses the higher degree nodes
first. For this reason, the disease spreading induces a second order
phase transition in the susceptible network with mean field exponents
at $t_c$ (see discussion in the Appendix Sec.II).

An important implication of our results is that, it can be used by the
health authorities to implement intervention strategies before the
critical time $t_c$ is reached. This will allow to protect
  a macroscopic fraction of the network composed by healthy interconnected
  individuals which preserve all the topological properties
  characteristic of social contact networks and their functionality.

\section*{Conclusions}\label{SecConc}
In this paper we introduce a temporal dynamic percolation to
characterize the evolution of the susceptible individuals in a SIR
model. We show using an edge-based compartmental approach and
percolation tools that as the disease spreads the evolution of the
susceptible network can be explained as a temporal node void
percolation that can be mapped instantaneously into static
percolation. We show that for transmissibilities above $T^{*}$, there
exist a critical time above which the giant susceptible cluster is
destroyed and the susceptible network overcomes a second order
transition with mean field exponents. All our theoretical results are
in excellent agreement with the simulations. Our findings are very
interesting from an epidemiological point of view since the existence
of a threshold time implies that when a very virulent disease reaches a
small number of susceptible individuals, the authorities have only a
limited time to intervene, in order to protect a big community
(susceptible giant component) that has not been already reached by the
epidemic, and to preserve the topological features of SF networks. Our
finding on the susceptible network could be extended to other
epidemics dynamics allowing to obtain a better description of the
effect of diseases spreading on social and technological networks.

\section{Acknowledgments}
The authors gratefully thanks to Erik M. Volz for a useful private
communication and to the anonymous reviewer for his/her deep reading
of our paper and his/her helpful comments. This work is part of a
research project of UNMdP and FONCyT (Pict 0293/2008).

\renewcommand{\baselinestretch}{1.5}
\renewcommand{\thefigure}{S\arabic{figure}}
\renewcommand{\theequation}{S\arabic{equation}}
\renewcommand{\thetable}{S\arabic{table}}
\renewcommand{\thesection}{\Roman{section}.}

\appendix
\setcounter{figure}{0}
\section{The EBCM approach}

\begin{table}[H]
\caption{Definitions} \renewcommand{\arraystretch}{1}
\begin{tabular}{lp{0.11cm}}  
\toprule[0.05cm]
\multicolumn{1}{c}{Variable/Parameter}&\multicolumn{1}{l}{Definition}\\\cmidrule{1-2}
\multicolumn{1}{c}{$\beta$}&\multicolumn{1}{l}{Infection rate or infection probability.}\\
\multicolumn{1}{c}{$\gamma$}&\multicolumn{1}{l}{Recovery rate.}\\
\multicolumn{1}{c}{$t_r$}&\multicolumn{1}{l}{Recovery time.}\\
\multicolumn{1}{c}{$T$}&\multicolumn{1}{l}{Transmissibility.}\\
\multicolumn{1}{c}{$\theta_t$}&\multicolumn{1}{p{12cm}}{Probability that a neighbor of a root node has not transmitted yet the disease to the root node at time $t$.}\\
\multicolumn{1}{c}{$\Phi_{S}(t)$}&\multicolumn{1}{p{12cm}}{Probability that a neighbor of a root node is susceptible at time $t$.}\\
\multicolumn{1}{c}{$\Phi_{I}(t)$}&\multicolumn{1}{p{12cm}}{Probability that an infected neighbor of a root node has not transmitted the disease to the root node at time $t$.}\\
\multicolumn{1}{c}{$\Phi_{R}(t)$}&\multicolumn{1}{p{12cm}}{Probability that a neighbor is recovered at time $t$  without having transmitted the disease to the root node.}\\
\bottomrule[0.05cm]
\end{tabular}
\label{tabla_casos}
\end{table}

In the EBCM approach, $\theta_t$ is the probability that a root node
has not being infected by a neighbor at time $t$. This is possible if
the neighbor is susceptible, recovered, or infected but has not
transmitted the disease yet to the root, which happens with
probabilities $\Phi_{S}(t)$, $\Phi_{R}(t)$ and $\Phi_{I}(t)$,
respectively. Then, $\theta_t=\Phi_{S}(t)+\Phi_{I}(t)+\Phi_{R}(t)$.
The probability that a root node of connectivity $k$ is susceptible is
$\theta^{k}$ and the fraction of susceptible nodes is $S(t)=\sum_{k}
P(k)\theta_t^{k}=G_{0}(\theta_t)$. On the other hand, a neighbor is
susceptible with probability $\Phi_{S}(t)=G_{1}(\theta_t)$. Then in
the SIR model with infection and recovery rates~\cite{Mil_03,Mil_04},
the probabilities $\Phi_{I}(t)$, $\Phi_{S}(t)$ and $\theta_t$ evolve
as,

\begin{eqnarray}
\dot{\theta}&=&-\beta\Phi_{I},\label{Eq_Mil_our_AppCon1}\\
\dot{\Phi}_{S}&=&-\beta G_{1}^{'}(\theta)\Phi_{I},\label{Eq_Mil_our_AppCon2}\\
\dot{\Phi}_{I}&=&-\beta\Phi_{I}+\beta G_{1}^{'}(\theta)\Phi_{I}-\gamma\Phi_{I},\label{Eq_Mil_our_AppCon3}
\end{eqnarray}
where $\beta$ and $\gamma$ are the infection and recovered
rates. Eq.~(\ref{Eq_Mil_our_AppCon1}) represents the decrease of
$\theta$ when an infected neighbor transmits the disease. The
Eq.~(\ref{Eq_Mil_our_AppCon2}) represents the decrease of $\Phi_{S}$
when a susceptible neighbor is infected, which is proportional to
$G_{1}^{'}(\theta)$, $i.e.$, the mean connectivity of the susceptible
first neighbors or the excess degree of the susceptible individuals,
because when a susceptible individual is infected, all its links
except the one used to infected it, can transmit the disease. This
term contributes to an increase of $\Phi_{I}$ in
Eq.~(\ref{Eq_Mil_our_AppCon3}). In Eq.~(\ref{Eq_Mil_our_AppCon3}) on
the $r.h.s$, the first term represents the decrease of $\Phi_{I}$ when
the links transmit the disease, the second term corresponds to the
term of Eq.~(\ref{Eq_Mil_our_AppCon2}) mentioned above and the third
term represents the decrease of $\Phi_{I}$ due to the recovery of
infected individuals.

To obtain the evolution of $I(t)$, we use the fact that,
\begin{eqnarray}
\dot{I}+\dot{S}+\dot{R}=0.
\end{eqnarray}
As $\dot{R}=\gamma\;I$ and
$\dot{S}=d\left(G_{0}(\theta)\right)/dt=G_{0}^{'}(\theta)\dot{\theta}=-\beta\Phi_{I}G_{0}^{'}(\theta)$,
the evolution of the fraction of infected individuals is given
by
\begin{eqnarray}
\dot{I}&=& \beta G_{0}^{'}(\theta)\Phi_{I}-\gamma I\label{Eq_Mil_our_AppCon4},
\end{eqnarray}
where the first term represents the decrease of $S$ which is
proportional to $\beta$, the mean connectivity of susceptible
individuals $G_{0}^{'}(\theta)$ and the probability that an outgoing
edge from a root is connected with an infected node that has not
transmitted the disease to the root at time $t$. The second term
corresponds to the recovery of infected individuals at a rate $\gamma$.

We reformulate the EBCM approach process with discrete time steps, for
a fixed recovery time $t_r$. It is straightforward that
Eq.~(\ref{Eq_Mil_our_AppCon1}-\ref{Eq_Mil_our_AppCon4}) can be written
as,
\begin{eqnarray}
\Delta{\theta_t}&=&-\beta\Phi_{I}(t),\label{Eq_Mil_ourAppe1}\\
\Delta{\Phi}_{S}(t)&=&G_{1}(\theta_{t+1})-G_{1}(\theta_{t}),\label{Eq_Mil_ourAppe2}\\
\Delta{\Phi}_{I}(t)&=&-\beta\Phi_{I}(t)-\Delta{\Phi}_{S}(t)+(1-T)\Delta{\Phi}_{S}(t-t_r),\label{Eq_Mil_ourAppe3}
\end{eqnarray}
where $1-T=(1-\beta)^{t_r}$ denotes the probability that an infected
individual has not transmitted the disease to a susceptible individual
during $t_r$ time units since he was infected. Finally the evolution
of the fraction of infected individuals is given by
\begin{eqnarray}
\Delta{I}(t)&=& -\Delta S(t)+\Delta S(t-t_r), \label{Eq_Mil_ourAppe4}
\end{eqnarray}
where $-\Delta
S(t)=-\left(G_{0}(\theta_{t+1})-G_{0}(\theta_{t})\right)$ represents
the fraction of new infected individuals and the second term
represents the recovery of infected individuals that have been
infected $t_r$ time units ago.

For the simulations we infect only one individual in the giant
component of the network and at each time step all the infected
individuals infect their susceptible network with probability $\beta$
and recover at a fixed time $t_r$ since they were infected. We select
only the runs in which the size of the epidemic has reached a
macroscopic fraction of individuals in the steady
state~\cite{Lag_02,New_05,Mil_02} because the deterministic equations
are only valid for epidemics above the critical threshold $T_c$. We
performed all the simulations using synchronized or simultaneous
updates at each time step.

\begin{figure}[H]
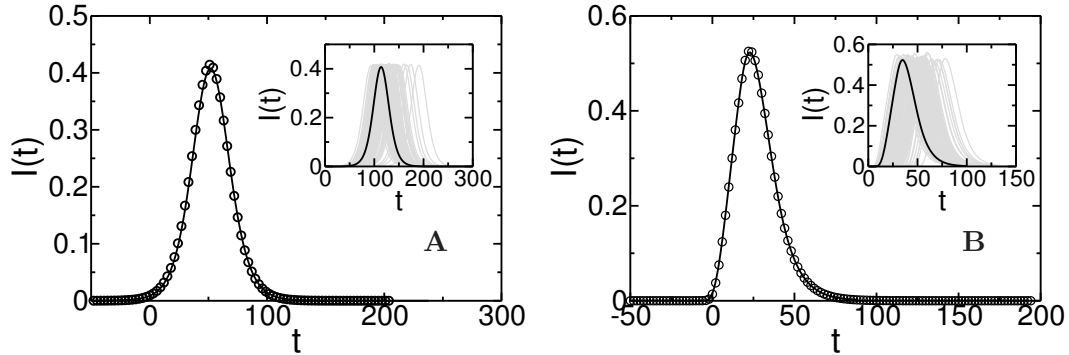

\centering
\vspace{1.0cm}
  \begin{overpic}[scale=0.25]{FigS1.eps}
    \put(80,20){{\bf{A}}}
  \end{overpic}\hspace{0.20cm}
  \begin{overpic}[scale=0.25]{FigS2.eps}
    \put(80,20){\bf{B}}
  \end{overpic}\vspace{0.25cm}
\caption{$I(t)$ for epidemics with $t_r=20$ and $\beta=0.04$
  ($T=0.55$) on networks with mean connectivity $4.07$ in the giant
  component, for a ER network with $\langle k \rangle=4$ (A) and a SF
  with $\lambda=2.63$, minimal connectivity $k_{min}=2$ and $\langle k
  \rangle =4.07$ (B). The symbols correspond to an average of one
  hundred different network realizations with $N=10^{5}$ nodes and the
  solid black curve is the numerical solution of
  Eq.~(\ref{Eq_Mil_ourAppe4}) shifting the curves to $t=0$ when $1$\%
  of the individuals are infected. The insets show the individual
  $100$ network realizations (solid gray lines) and the numerical
  solution of Eq.~(\ref{Eq_Mil_ourAppe4}) (solid black line) without
  the temporal shift transformation.}\label{EvInfERNew}
\end{figure}

In Fig.~\ref{EvInfERNew}, we plot the time evolution of the fraction
of infected nodes $I(t)$ for ER and SF networks obtaining by the EBCM
approach Eqs.~(\ref{Eq_Mil_ourAppe1}-\ref{Eq_Mil_ourAppe4}) and the
simulation. For the simulations we shifted $t=0$ to the instant when
the disease has reached $1$\% of the individuals. We choose this
reference time, as the time when the disease has reached a size enough
to growth deterministically. This choice compensates the time
dispersion of each trial around the theoretical solution due to
stochastic effects at the early stages of the process when the number
of infected nodes is small~\cite{Mil_04} (see the insets of
Fig.~\ref{EvInfERNew}).  As shown in Figs.~\ref{EvInfERNew}A-B,
each trial simulation has the same shape as the theoretical solution
which shows that the EBCM approach and the simulations are in
excellent agreement.

\section{Node void percolation in the time domain}

\begin{figure}[H]
\centering
\vspace{1cm}
 \includegraphics[scale=0.4]{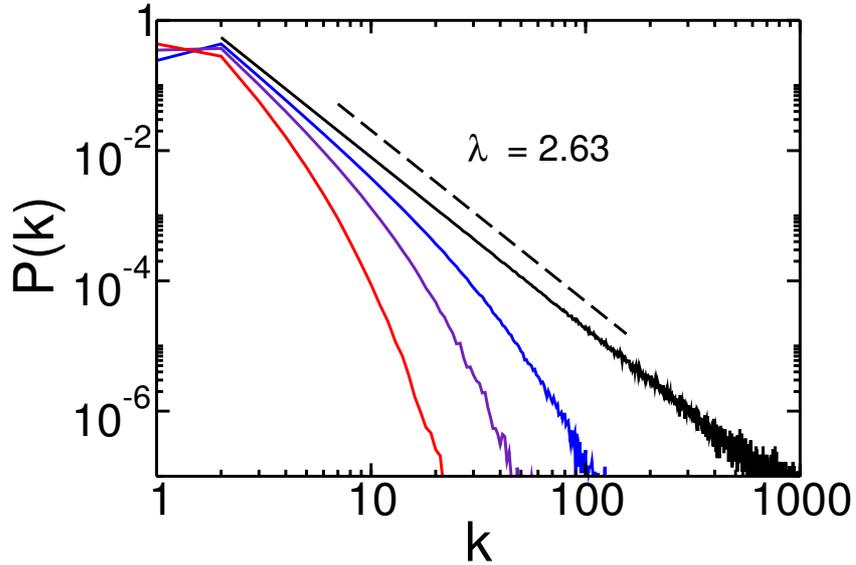}
  \vspace{0.5cm}
\caption{Simulation results of the degree distribution of susceptible
  nodes for a SF network with $\lambda=2.63$, $k_{min}=2$ and $\langle
  k \rangle =4.07$ at different times: at the beginning of the
  spreading (black solid line), when the disease has reached $10$\% of
  individuals (blue solid line), $25$\% of individuals (violet solid
  line) and $50$\% of individuals corresponding to $t_c$ (red solid
  line). (Color online). }\label{PkSFSus}
\end{figure}

\begin{figure}[H]
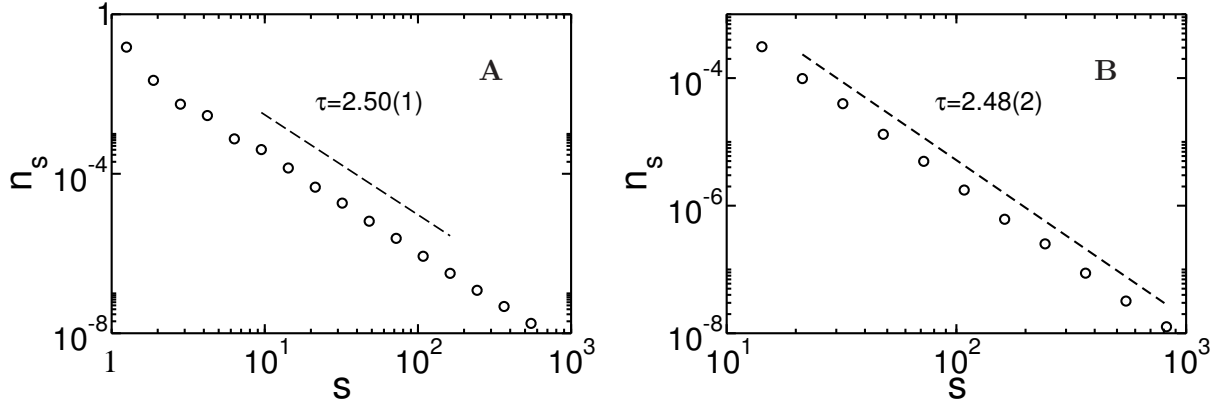

\centering
\vspace{1.0cm}
  \begin{overpic}[scale=0.28]{FigS4.eps}
    \put(80,55){{\bf{A}}}
  \end{overpic}\hspace{0.25cm}
  \begin{overpic}[scale=0.28]{FigS5.eps}
    \put(80,55){\bf{B}}
  \end{overpic}\vspace{0.8cm}
\caption{Log-log of the cluster size distribution $n_s$ of finite
  susceptible clusters ($\bigcirc$) at $t_c$ for $t_r=20$ and
  $\beta=0.07$ ($T$=0.76) in a ER network with $\langle k \rangle=4$
  for $t_c=30$(A) and a SF with $\lambda=2.63$, minimal connectivity
  $k_{min}=2$ and $\langle k \rangle =4.07$ for $t_c=11.$ (B). The
  dashed line corresponds to a power law fitting, from where we obtain
  an exponent $\tau \approx 2.5$. Our simulations were averaged over
  10000 network realizations with $N=10^{5}$.}\label{ns}
\end{figure}

In node void percolation, as a link is traversed, void node is
removed. The void nodes are removed with probability proportional to
$kP(k)$. As the susceptible nodes can be mapped into void node
percolation, the susceptible network loses their higher degree nodes
first as in an intentional attack. As a consequence, the resulting
susceptible network is more homogeneous than the original. Thus, mean
field exponents of a second order percolating phase
transition~\cite{Coh_04} are expected. In order to show the effect of the disease
spreading on the highest degree nodes, in Fig.~\ref{PkSFSus} we plot
for a SF network the effective degree distribution of the susceptible
nodes obtained from the simulations, in which a susceptible node has
degree $k$ when it has $k$ susceptible neighbors.
As shown in Fig.~\ref{PkSFSus}, as the disease spreads, the effective
degree distribution loses the heavy tail. As a result of this process,
the susceptible clusters becomes more sparse and at the critical time
$t_c$ the topology of the susceptible clusters change drastically
since the susceptible individuals lose all the hubs and $P(k)$ has an
exponential tail. For percolation in mean field it is known that at
the criticality the finite cluster size distribution $n_S\sim
s^{-\tau}$ with $\tau=2.5$ and $S_1\left[\Phi_s(t)\right]\sim
\Phi_s(t)-\Phi_s(t_c)$. In Fig.~\ref{ns} we plot the simulations
results of the finite size distribution of the susceptible nodes $n_s$
at $t=t_{c}$.
\begin{figure}[H]
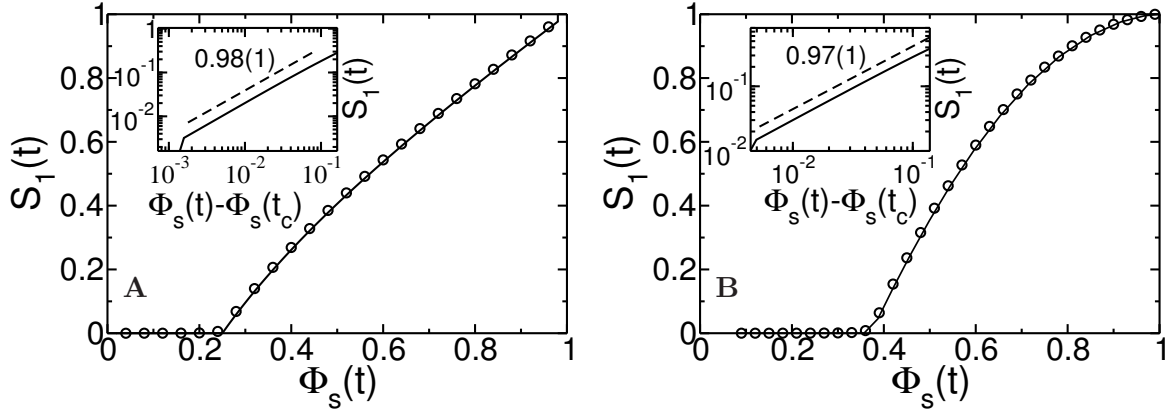

\centering
\vspace{1.0cm}
  \begin{overpic}[scale=0.28]{FigS6.eps}
    \put(20,20){{\bf{A}}}
  \end{overpic}\hspace{0.25cm}
  \begin{overpic}[scale=0.28]{FigS7.eps}
    \put(20,20){\bf{B}}
  \end{overpic}\vspace{0.8cm}
\caption{$S_1$ as a function of $\Phi_s(t)$ obtained from simulations
  ($\bigcirc$) and from the analytical approach (solid line) in a ER
  network with $\langle k \rangle=4$ (A) and a SF with $\lambda=2.63$,
  minimal connectivity $k_{min}=2$ and $\langle k \rangle =4.07$
  (B). In the inset we plot $S_1$ as a function of the distance of
  $\Phi_s(t)$ to the criticality $\Phi_{s}(t_c)=V_c$, in log-log
  scale. The dashed line corresponds to a power law fitting from where
  we obtain slope $\sim 1$. Our simulations were averaged over 1000
  network realizations with $N=10^{5}$.}\label{SmedCrit}
\end{figure}
We can see that at $t_c$, $n_s(t_c)$ behaves as a power law with
exponent $\tau\approx 2.5$ which corresponds to the mean field value,
independently of the initial degree distribution of the
network~\cite{Exxon_07}. Similarly, in Fig.~\ref{SmedCrit} we plot
$S_1(t)$ as a function of $\Phi_S(t)$ obtained from the simulations
and the theoretical approach.  We compute $\Phi_S(t)$ from
  the simulations as the square root of the fraction of edges
  connecting two susceptible nodes~\cite{Exxon_06}. We can see that
$S_1(t)$ behaves as a power law with exponent one with the distance to
the criticality $\Phi_s(t_c)$, which also corresponds to the mean
field value (see Insets of Fig.~\ref{SmedCrit}). Since two critical
exponents are sufficient to determine the universality class, the
results showed above indicate that in a node void percolation process
the susceptible network belongs to the same universality class of mean
field percolation and confirms quantitatively the homogenization of
the susceptible network during a SIR epidemic spreading.

\begin{figure}[H]
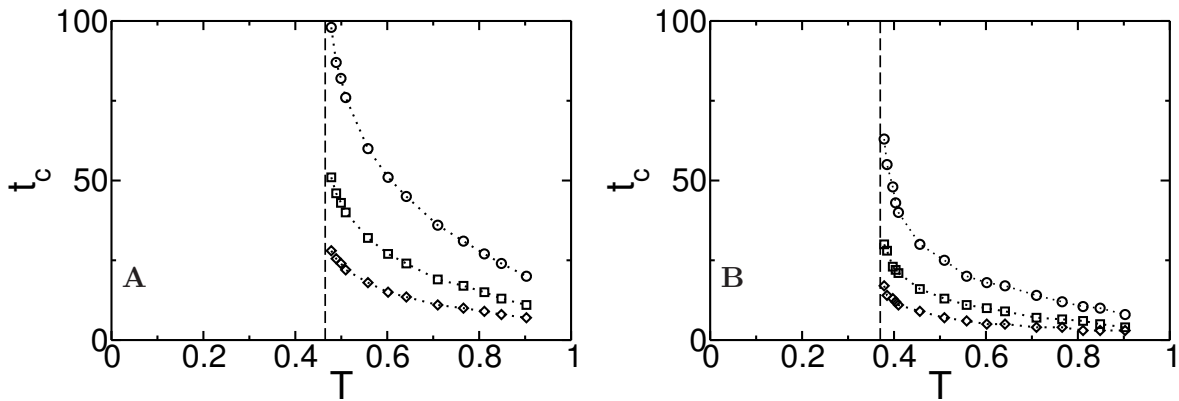

\centering
\vspace{1.0cm}
  \begin{overpic}[scale=0.28]{FigS8.eps}
    \put(20,20){{\bf{A}}}
  \end{overpic}\hspace{0.25cm}
  \begin{overpic}[scale=0.28]{FigS9.eps}
    \put(20,20){\bf{B}}
  \end{overpic}\vspace{0.25cm}
\caption{$t_c$ as a function of $T$ for $\beta=0.07$ and $t_r=20$
  ($\bigcirc$), $t_r=10$ ($\square$), $t_r=5$ ($\diamond$) and mean
  connectivity $4.07$ in the giant component in a ER network with
  $\langle k \rangle=4$ ($T^{*}=0.46$) (A) and in a SF with
  $\lambda=2.63$, minimal connectivity $k_{min}=2$ and $\langle k
  \rangle =4.07$ ($T^{*}=0.38$)(B). The dashed line represents the
  value of $T^{*}$. The critical time $t_c$ is measured using $t=0$
  when $1$\% of individuals are infected. The dotted lines are used as
  a guide to the eyes.}\label{Critical_Time}
\end{figure}

Finally, in Fig.~\ref{Critical_Time}, we plot the critical time
$t_{c}$, computed from the simulations at $\Phi_S(t)=V^{s}_c$, as a
function of $T$ for different values of $t_r$. We can see that for
the same transmissibility $T$, when $t_r$ increase, the time to intervene
grows since $\beta$ decrease and thus the disease spreading is
retarded. In turn, when the transmissibility $T$ reaches $T^{*}$ from
above, the critical time $t_c$ grows very fast. This phenomenon is analogous to
other second order phase transitions in physics like the relaxation
time near the Curie temperature, which are called ``critical slowing
down''~\cite{Sta_01,Bun_01}, and indicates that that once the
transmissibility increases slightly above $T^{*}$, the time needed to
destroy the giant susceptible cluster decreases very fast.

\bibliography{bibpaper_corr_sup}

\begin{thebibliography}{26}
\bibitem[{Exx({\natexlab{a}})}]{Exxon_07}
\bibinfo{note}{For SF networks with $2<\lambda<4$,
  $\tau=(2\lambda-3)/(\lambda-2)$ and the exponent of the order parameter is
  $1/(3-\lambda)$ for $2<\lambda<3$, $1/(\lambda-3)$ for $3<\lambda<4$ and one
  for $\lambda>4$}.
\bibitem[{Exx({\natexlab{b}})}]{Exxon_06}
\bibinfo{note}{As $\Phi_S(t)$ is the probability that a randomly chosen stub
  belongs to a susceptible node (conditional on the assumption that stub has
  not transmitted infection to the node) then the probability that both stubs
  in a random edge belong to susceptible nodes is $\left(\Phi_S(t)\right)^2$.}
\end{thebibliography}

\end{document}